\pdfoutput=1

\documentclass[11pt]{article}

\usepackage[preprint]{acl}

\usepackage{algorithm}
\usepackage{algpseudocode}

\usepackage{times}
\usepackage{latexsym}
\usepackage{url}

\usepackage{todonotes}

\usepackage[T1]{fontenc}

\usepackage[utf8]{inputenc}

\usepackage{microtype}

\usepackage{inconsolata}

\usepackage{amsmath} 
\usepackage{amssymb}
\usepackage{array} 
\usepackage{multirow}
\usepackage{graphicx}
\usepackage{microtype}
\usepackage{inconsolata}

\usepackage{times}
\usepackage{latexsym}
\usepackage{adjustbox} 
\usepackage{graphicx}
\usepackage{listings}
\usepackage{booktabs}
\usepackage{hyperref}

\usepackage{url}

\usepackage{xcolor}
\usepackage{enumitem}
\PassOptionsToPackage{dvipsnames, svgnames, x11names}{xcolor}
\usepackage{tcolorbox} 

\title{Aligning NLP Models with Target Population Perspectives using PAIR: \\Population-Aligned Instance Replication}

\author{
  \textbf{Stephanie Eckman\textsuperscript{$\ast$,$\clubsuit$}}~~~
    \textbf{Bolei Ma\textsuperscript{$\ast$,$\heartsuit$}}~~~
  \textbf{Christoph Kern\textsuperscript{$\heartsuit$}}~~~
  \textbf{Rob Chew\textsuperscript{$\spadesuit$}}~~~
\\
  \textbf{Barbara Plank\textsuperscript{$\heartsuit$}}~~~
  \textbf{Frauke Kreuter\textsuperscript{$\clubsuit$,$\heartsuit$}}
\vspace{4.5pt}
\\
\textsuperscript{$\clubsuit$}University of Maryland, College Park~~~\\
\textsuperscript{$\heartsuit$}LMU Munich \& Munich Center for Machine Learning~~~\\
\textsuperscript{$\spadesuit$}RTI International~~~
\vspace{2.5pt}
\\
\small{$^\ast$Equal contributions.}
\vspace{1.5pt}\\
\small{\texttt{\{steph, fkreuter\}@umd.edu, \{bolei.ma, christoph.kern, b.plank\}@lmu.de, rchew@rti.org}}
}

\begin{document}
\maketitle

\begin{abstract}
Models trained on crowdsourced annotations may not reflect population views, if those who work as annotators do not represent the broader population. In this paper, we propose \textbf{PAIR}: \textbf{P}opulation-\textbf{A}ligned \textbf{I}nstance \textbf{R}eplication, a post-processing method that adjusts training data to better reflect target population characteristics without collecting additional annotations. Using simulation studies on offensive language and hate speech detection with varying annotator compositions, we show that non-representative pools degrade model calibration while leaving accuracy largely unchanged. PAIR corrects these calibration problems by replicating annotations from underrepresented annotator groups to match population proportions. We conclude with recommendations for improving the representativity of training data and model performance.\footnote{The code for experiments is available at \url{https://github.com/soda-lmu/PAIR}.}
\end{abstract}

\section{Introduction and Inspiration}
When a hate speech detection model flags harmless expressions as toxic, or a content moderation system fails to identify genuinely harmful content, the root cause often lies not in the model architecture, but in who annotated the training data. While Natural Language Processing (NLP) models aim to serve broad populations, the human judgments used to train these systems often come from crowdworkers and convenience samples. And the demographics, cultural contexts, and worldviews of these annotators often differ from those of the communities the models ultimately impact \cite{sorensen2024position, Fleisigetal05}. These non-representative annotator pools can have real consequences, because annotator characteristics like age, education level, and cultural background impact how content is annotated \cite{sap2022annotators, Fleisig2023WhenTM, Kirk_PRISM}. 
The influence of annotator characteristics underscores that language understanding is not a single objective truth but a constellation of equally valid interpretations anchored in different lived experiences. When this perspectivist interpretation is ignored, models trained on non-representative data can perpetuate the biases and blind spots of their limited training data \cite{Berinsky_Huber_Lenz_2012, multicalibration, mehrabi_survey_2021, rolf2021representation, hullermeier2021aleatoric, ouyang2022traininglanguagemodelsfollow, favier2023fair, smart2024disciplinelabelweirdgenealogy}.

\begin{figure}
    \centering
    \includegraphics[width=0.999\linewidth]{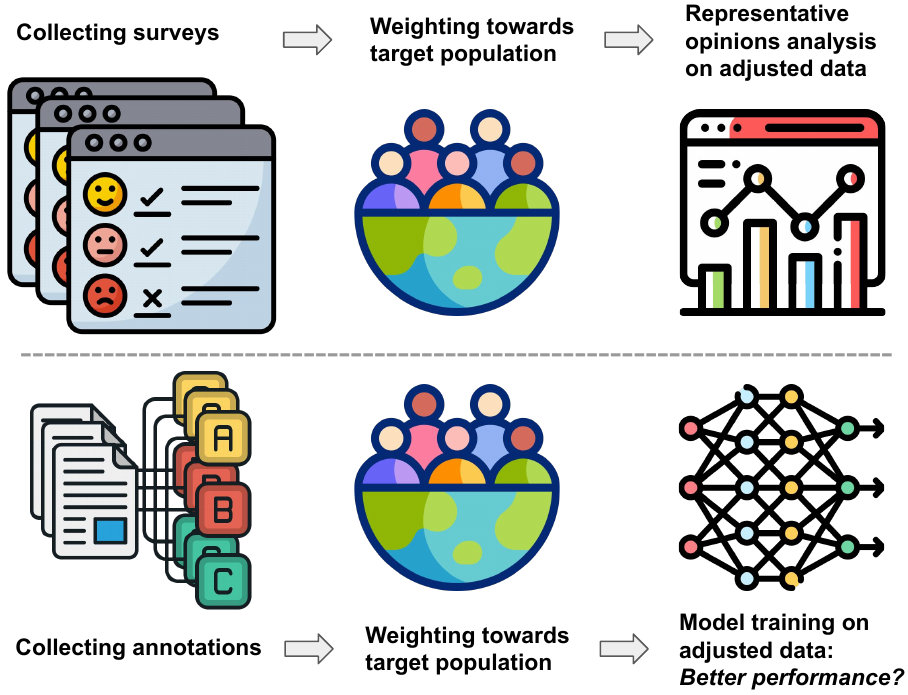}
    \caption{\textbf{Top}: Adjusting survey data to match population produces high quality results. \\\textbf{Bottom}: Can a similar adjustment in data annotations also improve model performance?}
    \label{fig:weighting}
\end{figure}

Fortunately, survey researchers have developed statistical techniques to produce population-level estimates from non-representative samples \cite{Bethlehem2011}. The top panel of Figure \ref{fig:weighting} shows a simple survey workflow: collecting survey data, creating statistical weights to match the sample to the population, and estimating population parameters. We adapt this approach to the machine learning context, enabling models to better align with target populations even when trained on non-representative annotator pools (bottom panel).

Our Population-Aligned Instance Replication (PAIR) method post-processes training data to better reflect target populations without collecting additional annotations. We test the approach with a simulation study \citep{Burton2006,Valliant2019,Morris2019} and answer two questions: 

\begin{itemize}
\item \textbf{RQ1:} How do non-representative annotator pools impact model calibration and accuracy?
\item \textbf{RQ2:} Can our proposed weighting method (PAIR) mitigate these annotator pool effects?
\end{itemize}

Our results demonstrate that models trained on non-representative annotator pools perform worse than those trained on representative pools. However, simple adjustment methods can improve performance without collecting additional data. These findings suggest that insights from survey methodology can help artificial intelligence (AI) systems better represent the populations they serve.

\section{Related Work}

Several strands of related work inform our approach to identifying and mitigating bias due to the use of non-representative annotators:

\paragraph{Annotator Impact on Data and Models.} Annotator characteristics and attitudes significantly influence annotation quality, particularly for subjective tasks like toxicity detection \citep{giorgi2024human, Prabhakaran2021, Fleisig2023WhenTM, sap2022annotators}. For example, annotators' political views and racial attitudes affect their toxicity judgments \citep{sap2022annotators}. Models trained on non-representative annotator pools inherit these biases and generalize poorly \citep{Berinsky_Huber_Lenz_2012, mehrabi_survey_2021, rolf2021representation, ouyang2022traininglanguagemodelsfollow,favier2023fair,smart2024disciplinelabelweirdgenealogy,mokhberian-etal-2024-capturing}. 

\paragraph{Annotator Demographics.} Several researchers advocate collecting annotator demographics to assess representation and identify biases \citep{bender2018data, Prabhakaran2021, plank-2022-problem, Wan2023,santy-etal-2023-nlpositionality,pei-jurgens-2023-annotator}.\footnote{In our context, these characteristics are used only to analyze bias. Because they are not available for unannotated text, they are not features that the model can use.} However, collecting and releasing these data can raise privacy concerns \citep{Fleisig2023WhenTM}. Recent works have also used demographics to prompt the large language models \cite{argyle2023out}, and some find that these are less effective in subjective contexts \cite{sun-etal-2025-sociodemographic,orlikowski2025demographics}.

\paragraph{Debiasing \& Data Augmentation Methods.} Prior work has proposed various approaches to reduce bias in training data features and annotations. Most similar to our work is the resampling and reweighting approaches of \citet{Caldersetal2009} and \citet{kamiran2012data}, imputation \cite{LowmanstoneWOKK23},  and the oversampling of minority class cases of \citet{LingDirectMarketing}. PAIR adapts these methods to balance \textit{annotator} characteristics rather than class labels or sensitive observation\--level features. PAIR retains the simplicity and interpretability of earlier resampling methods while extending them to a ``Learning with Disagreement'' \citep{uma2021learning, leonardelli2023semeval2023task11learning} setting with multiple annotations per observation, by replicating annotations from underrepresented annotator groups.

\section{PAIR Algorithm: Adjustment via Pseudo-Population}
\label{sec:adjust_details}

To adjust an annotator pool to better reflect a target population, we propose the PAIR algorithm, which constructs a pseudo-population through post-stratification, weight normalization, and deterministic replication. This adjustment strategy is inspired by established methods in survey sampling \citep{quatember-2015}.

Post-stratification aligns a sample more closely with population-level distributions \citep{Bethlehem2011, valliant2013practical}. Annotators are grouped into strata based on demographic or behavioral characteristics. For each unit $i$ in stratum $s$, a post-stratification weight is computed as:

\begin{equation}  \label{equ:pair}
w_{s,i} = \frac{P_s}{S_s}
\end{equation}

\noindent where $P_s$ and $S_s$ denote the share of the population in stratum $s$ and the share of the sample (or annotator pool), respectively. The $P$ values come from official statistics or surveys. The $S$ values likely come from the annotators themselves and researchers may have to collect them. This technique can accommodate multiple stratification variables; it is only limited by the availability of population or reference data and data about the annotators.

These weights have only relative meaning and are invariant to multiplication by a constant ($K$): 
\begin{equation} 
w_i^{\text{normalized}} = w_i^{\text{initial}} \times K
\end{equation} 
\noindent Normalization useful if research teams have a target number of annotations per observation in mind, for either computational or design reasons, or if some weights given by Eq. \ref{equ:pair} are very small and round to one.

To generate a pseudo-population, we apply deterministic replication: each unit is replicated $n_i$ times where
\begin{equation} 
n_i = \text{round}(w_i^{\text{normalized}}) - 1
\end{equation} 
ensuring integer replication counts. This approach produces a dataset that reflects population proportions while maintaining interpretability and reproducibility.

While we focus in this initial study on deterministic replication, alternative implementations are possible, including resampling-based replication or direct incorporation of weights into model training.

\section{Annotation Simulation and Model Training}
\label{sec:methods}

To address our research questions, we conduct a simulation study on offensive language and hate speech detection. We imagine a population made up of equal shares of two types of people: those more likely to perceive offensive language and hate speech and those less likely. We create three datasets of simulated annotations which differ in the mix of the annotator types. We then create a fourth dataset, using the PAIR algorithm, to fix the imbalance in the annotators. We fine-tune RoBERTa models on the four datasets and evaluate the effect of annotator composition on model performance (RQ1) and the ability of the PAIR algorithm to improve performance (RQ2). 

\subsection{Simulating Annotations}

\begin{figure}[htb]
    \centering
    \includegraphics[width=1\linewidth]{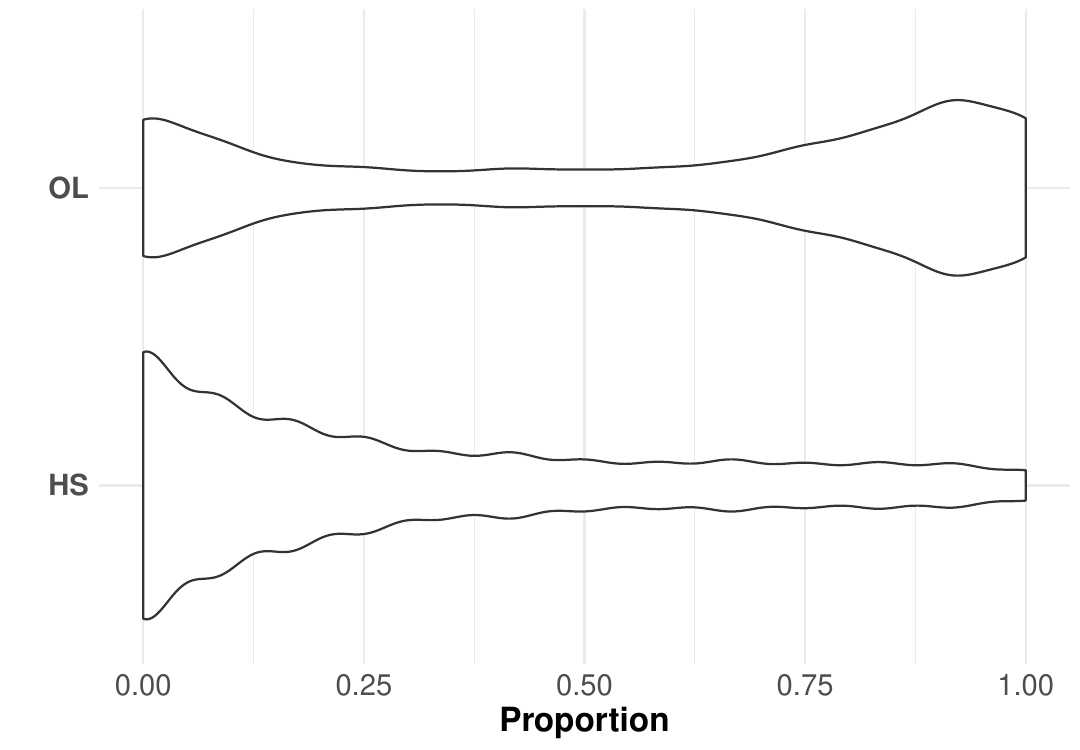}
    \caption{Distribution of $p_{i,OL}$ and $p_{i,HS}$ in original data}
    \label{fig:gold_dist}
\end{figure}

We use our previously collected dataset on tweet annotation sensitivity \citep{kern-etal-2023-annotation}\footnote{\url{https://huggingface.co/datasets/soda-lmu/tweet-annotation-sensitivity-2}}, which is a dataset of 3,000 English-language tweets, each with 15 annotations of both offensive language (OL: yes/no) and hate speech (HS: yes/no). We chose this dataset because the high number of annotations of each tweet gives us a diverse set of labels to work with. We randomly select (without replacement) 12 annotations (of both OL and HS) of each tweet in the original dataset.\footnote{As shown in Table \ref{tbl:datasets}, we can more carefully control the construction of our datasets when the number of annotations per tweet is even.} Let $p_{i,OL}$ be the proportion of the 12 annotators who annotated tweet $i$ as OL and $p_{i,HS}$ defined similarly. Figure \ref{fig:gold_dist} shows the distribution of these proportions across the 3,000 tweets. The HS annotations are clustered near 0, whereas the OL annotations are more spread out between $0$ and $1$. 

The population contains two types of people (50\% each). \textbf{Type A} people are \textit{less likely} to say a tweet contains OL. \textbf{Type B} people are \textit{more likely}:

\begin{align}
p_{i,OL}^{A} &= \text{max}(p_{i,OL} - \beta, 0) \\
p_{i,OL}^{B} &= \text{min}(p_{i,OL} + \beta, 1) 
\label{eq:p_norm}
\end{align}

Here $\beta$ captures the magnitude of the bias. We vary $\beta$ from $[0.05, 0.3]$ by $0.05$, corresponding to an increase or decrease in the probability to judge a tweet as OL by five to 30 percentage points. This range is large on the probability scale and covers most reasonable situations. With these six values of $\beta$, we create six vectors of probabilities ($p_{i,OL}^A, p_{i,OL}^B$) for each tweet.

We then create four datasets, each with 3,000 tweets (Table \ref{tbl:datasets}), for each value of $\beta$. The \textbf{Representative} Dataset contains OL annotations from six A annotators (drawn from $\text{Bernoulli}(p_{i,OL}^{A})$) and six B annotators (drawn from $\text{Bernoulli}(p_{i,OL}^{B})$). The proportion of A and B annotators in this dataset matches the simulated population we created. 

\begin{table}[ht]
  \small
  \centering
  \renewcommand\arraystretch{1.15}
  \setlength\tabcolsep{2pt}
    \begin{tabular}{lccc}
    \toprule
 & Annotations & A & B   \\
Dataset& per tweet& annotations  & annotations \\
\midrule
Representative & 12 & 6 & 6 \\
Non-representative 1 &  9 & 6 & 3 \\
Non-representative 2 & 12 & 9 & 3 \\
Adjusted & 12 & 6 & 3 + 3* \\
\bottomrule
\multicolumn{4}{l}{* 3 B annotations replicated} \\
    \end{tabular}
    \caption{Four training datasets for each bias value ($\beta$)}
    \label{tbl:datasets}
\end{table}

We next create two unbalanced datasets. \textbf{Non-representative 1} randomly deletes three B annotations for each tweet from the Representative Dataset. \textbf{Non-representative 2} adds three additional A annotations, drawn from $p_{i,OL}^A$, to the Non-representative 1 dataset. The Non-representative 2 Dataset is more unbalanced than Non-representative 1, but contains the same number of annotations as the Representative dataset. 

\subsection{Applying PAIR Algorithm}

Finally, we use the PAIR algorithm to create the \textbf{Adjusted} Dataset. Starting with the Non-representative 1 Dataset, we calculate the share of the annotator pool that is in the A and B strata: $S_A = \frac 2 3, S_B = \frac 1 3$. The population proportions, by construction, are $P_A = 0.5, P_B = 0.5$. Applying \eqref{equ:pair}, we get $w_{A, i} = 0.75, w_{B,i}=1.5$. We multiply these weights by $K=\frac 4 3$ to get $w_{A, i} = 1, w_{B,i}=2$. These weights give us $n_{A,i} = 0, n_{B,i}=1$, which leads us to replicate all B annotations in the Non-representative 1 Dataset (see Table \ref{tbl:datasets}).

\smallskip
The HS probabilities for the A and B annotators are defined in the same way: 
$p_{i,HS}^{A} = \text{max}(p_{i,HS} - \beta, 0)$, 
$p_{i,HS}^{B} = \text{min}(p_{i,HS} + \beta, 1)$. We also construct the four datasets (Representative, Non-representative 1, Non-representative 2, Adjusted) in the same way we did in the OL case.

\begin{figure}[htbp]
    \centering
    \includegraphics[width=1\linewidth]{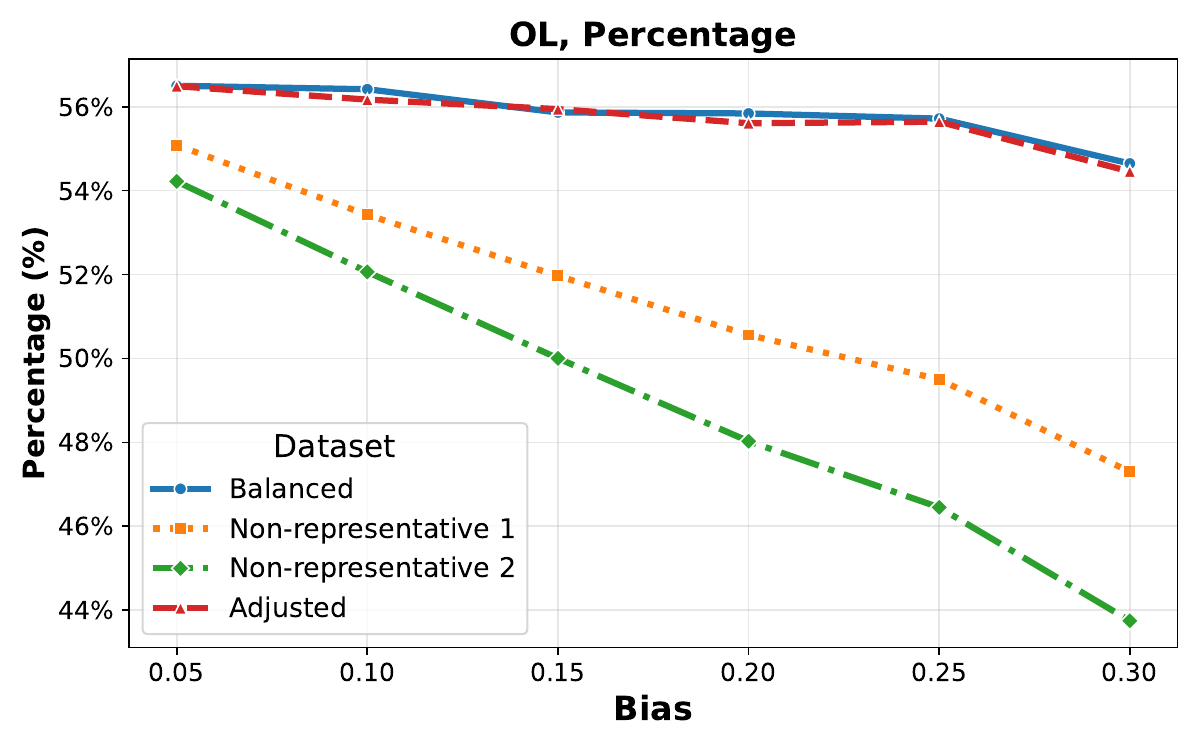}
    \caption{Percentage of instances annotated as OL, by dataset and bias ($\beta$)}
    \label{fig:percentage_ol}
\end{figure}

\begin{figure}[htbp]
    \centering
    \includegraphics[width=1\linewidth]{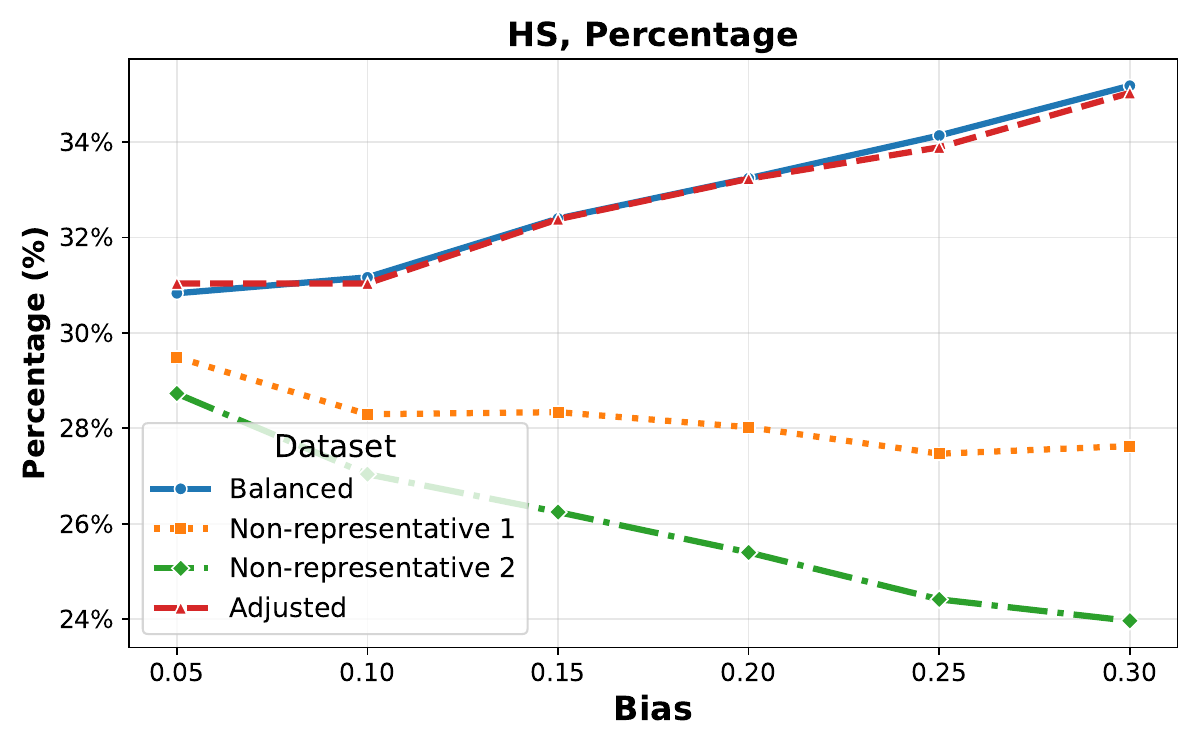}
    \caption{Percentage of instances annotated as HS, by dataset and bias ($\beta$)}
    \label{fig:percentage_hs}
\end{figure}

Figures \ref{fig:percentage_ol} and \ref{fig:percentage_hs} show the percentage of instances annotated OL and HS in the four datasets for each value of $\beta$. In both, the percentage of OL/HS annotations in the Adjusted dataset is similar to that in the Representative dataset for all values of $\beta$. The percentage in the two unbalanced datasets is lower, because those datasets overrepresent the A annotators, who are less likely to annotate OL/HS.

HS is rare in our dataset (16.7\% of instances were annotated as HS), and our simulation strategy overrepresents A annotators in the two Non-representative datasets, who are less likely to perceive HS (Table \ref{tbl:datasets}). For these reasons, as $\beta$ increases, more  $p_{i,HS}^A$ are $0$ while the $p_{i,HS}^B$ probabilities increase. This issue leads the proportion of HS annotations in the Representative and Adjusted datasets to increase with $\beta$ in the HS dataset, which have more B annotations than the unadjusted datasets (Figure \ref{fig:percentage_hs}).

\subsection{Model Training and Evaluation}

\paragraph{Training and Test Setup.} We train models on each of the eight datasets: four for OL, four for HS. We divide each dataset, at the tweet level, into training (2,000 tweets), development (500), and test (500) sets. Each tweet appears 12 times in the Representative, Non-representative 2, and Adjusted datasets and nine times in the Non-representative 1 set.

\paragraph{Model Selection and Training.} 
We used RoBERTa base \cite{liu2019roberta} as our text classifier, training for five epochs on each dataset, with development set optimization. To ensure reliable results, we trained five versions with different random seeds and averaged their performance. 


Our implementation of RoBERTa models was based on the libraries \texttt{pytorch} \citep{pytorch2019} and \texttt{transformers} \citep{wolf-etal-2020-transformers}. During training, we used the same hyperparameter settings (Table \ref{hyperparambert}) for the five training conditions to keep these variables consistent for comparison purposes. We trained each model variation with five random seeds $\{10, 42, 512, 1010, 3344\}$ and took the average across the models. All experiments were conducted on an NVIDIA$^\circledR$ A100 80 GB RAM GPU. 

\begin{table}[htbp]
  \centering
  \small
  \renewcommand\arraystretch{1.15}
  \begin{tabular}{lr}
\toprule
    Hyperparameter & Value\\
    \midrule
    encoder & \texttt{roberta-base} \\
    epochs\_trained &    $5$ \\
    learning\_rate   & $3e^{-5}$ \\
    batch\_size & $32$ \\
    warmup\_steps & $500$ \\
    optimizer & \texttt{AdamW} \\
    max\_length & $128$ \\
\bottomrule
  \end{tabular}
\caption{Hyperparameter settings of RoBERTa models}
\label{hyperparambert}
\end{table}

\paragraph{Performance Metrics.}
We evaluate models using \textbf{calibration and accuracy metrics} on the test set. While accuracy metrics directly measure classification performance, calibration metrics provide crucial insights into model reliability by assessing probability estimate quality -- particularly important for high-stakes applications requiring trustworthy confidence measures.

For \textbf{calibration}, we report Absolute Calibration Bias (ACB, Equation \ref{equ:acb}), which measures how well a model's predicted probabilities align with true annotation frequencies. For each tweet $i$, we compare the model's predicted probability of offensive language ($\text{preds}_{i,OL}$) against the true proportion of annotators who labeled that tweet as offensive ($p_{i,OL}$).

\begin{align}\label{equ:acb}
\text{ACB}_{OL} =  \frac{1}{n} \sum_{i=1}^n \left| \text{preds}_{i,OL} - p_{i,OL}\right|
\end{align}

$\text{ACB}_{HS}$ is defined accordingly. ACB adapts established calibration metrics by using the annotator agreement proportion as a plug-in estimator for the true probability, avoiding the need for binning (as in ECE, \citealp{Naeini_Cooper_Hauskrecht_2015}) while maintaining the intuitive L1 distance interpretation \citep{pmlr-v151-roelofs22a}. A low ACB score indicates that the model's confidence scores accurately reflect the underlying annotation uncertainty in the population.

For \textbf{accuracy}, we report the F1 score. 

\section{Results}\label{sec:results}

We show results separately for the OL and HS models.

\subsection{OL Models}\label{sec:ol_results}

\begin{figure}[t]
    \centering
    \includegraphics[width=1\linewidth]{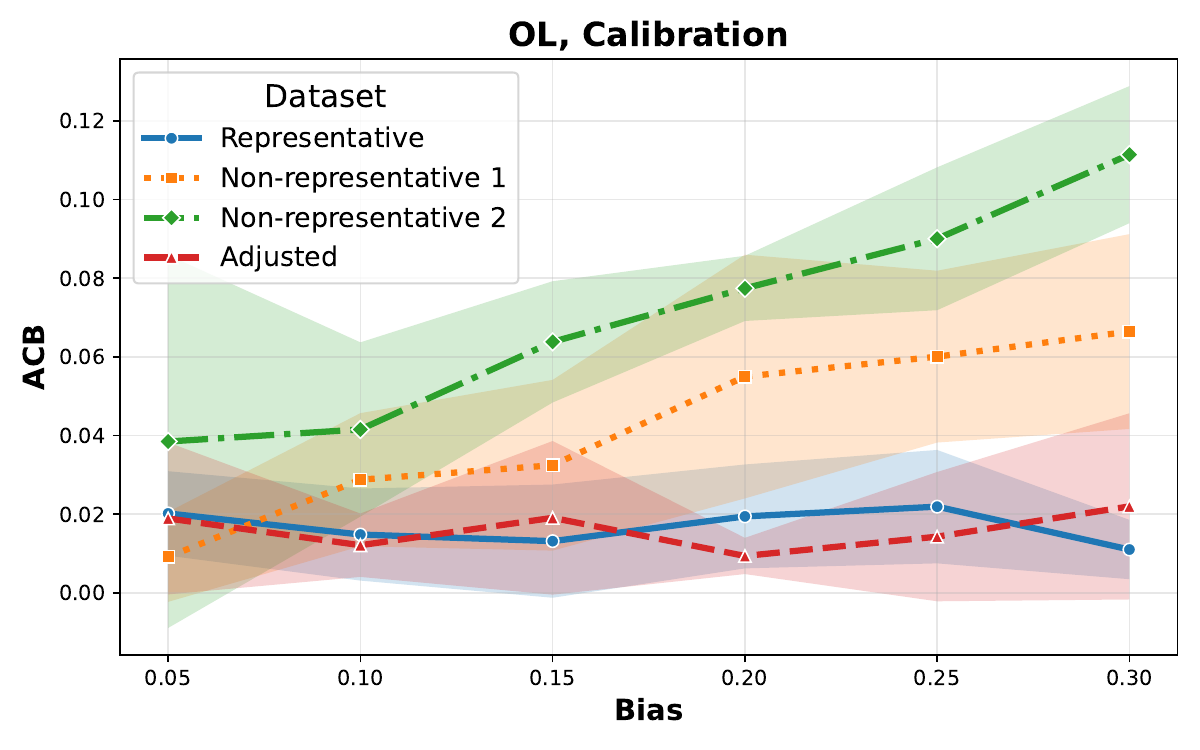}
    \caption{ACB scores for OL Models, by dataset and bias ($\beta$)}
    \label{fig:acb-ol}
\end{figure}

\paragraph{Calibration.} Figure \ref{fig:acb-ol} compares the ACB in the test set for models trained on the four datasets. The dark lines show average ACB across the five training runs and the shading shows the standard deviation.

The ACB for the models trained on the Adjusted dataset closely tracks that for the Representative dataset and does not increase with $\beta$. ACB for the models trained on the two unbalanced datasets is greater and grows with $\beta$. These results demonstrate the effectiveness of our adjustment method. Replicating the annotations from the underrepresented annotator type to match population proportions improves model calibration.

\begin{figure}[t]
    \centering
    \includegraphics[width=1\linewidth]{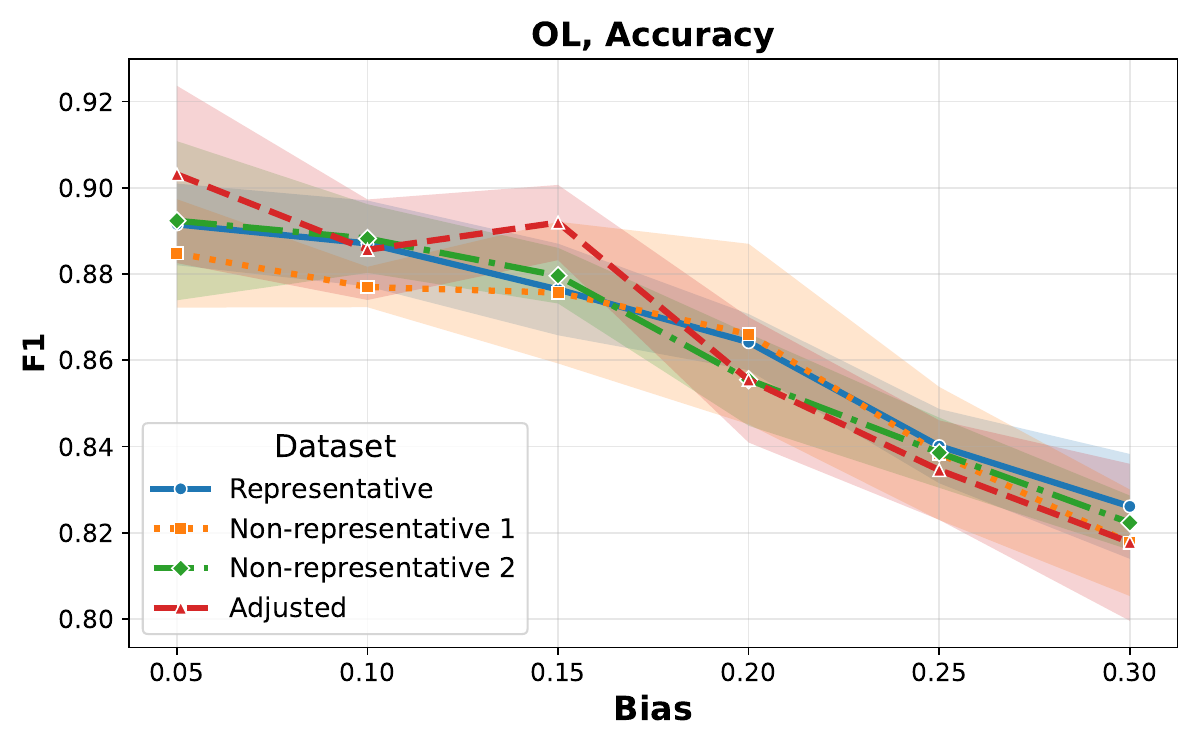}
    \caption{F1 scores for OL Models, by dataset and bias ($\beta$)}
    \label{fig:f1-ol}
\end{figure}

\begin{figure}[htbp]
    \centering
    \includegraphics[width=1\linewidth]{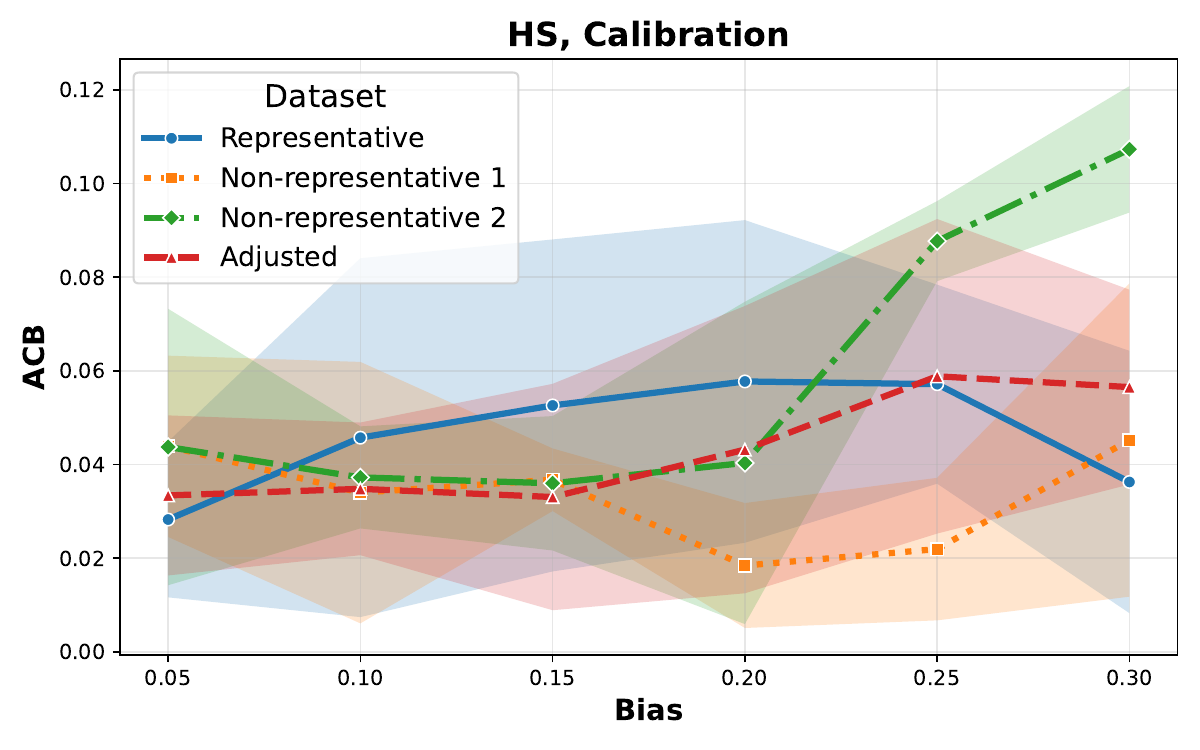}
    \caption{ACB scores for HS Models, by dataset and bias ($\beta$)}
    \label{fig:acb-hs}
\end{figure}

\begin{figure}[htbp]
    \centering
    \includegraphics[width=1\linewidth]{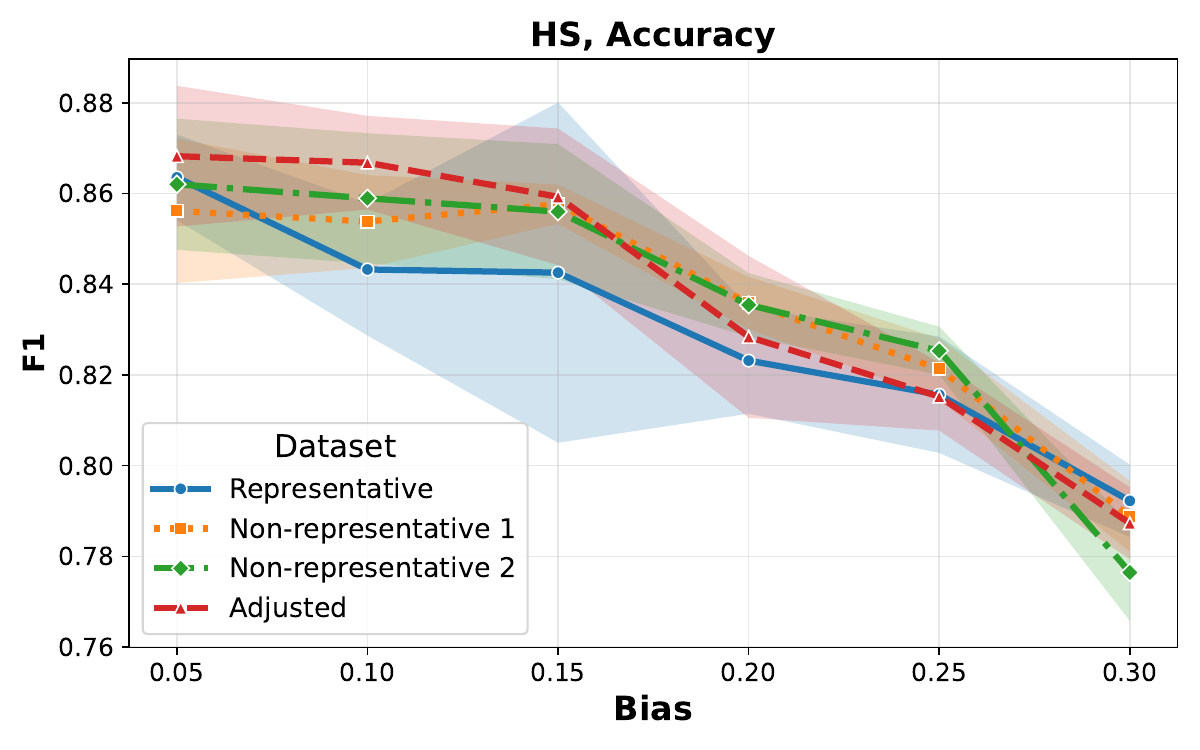}
    \caption{F1 scores for HS Models, by dataset and bias ($\beta$)}
    \label{fig:f1-hs}
\end{figure}

\begin{figure}[t]
    \centering
    \includegraphics[width=1\linewidth]{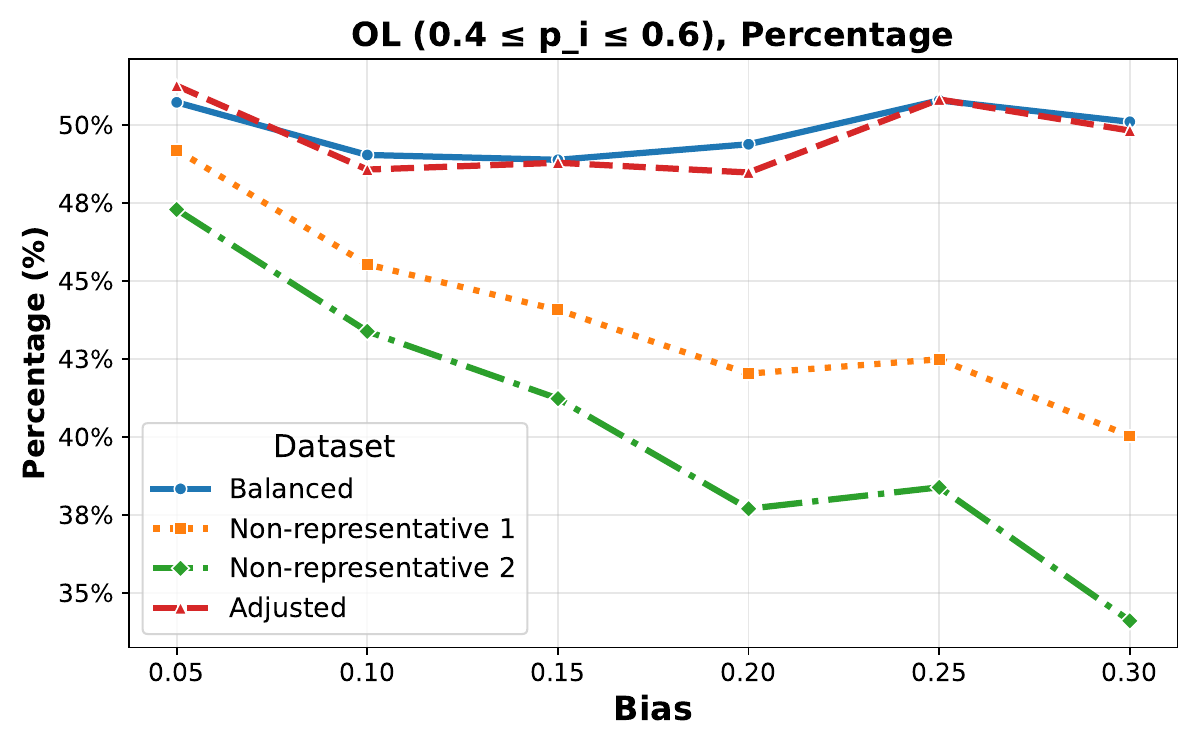}
    \caption{Percentage of OL instances on difficult tweets ($0.4 \leq p_{i,OL} \leq 0.6$) by dataset and bias ($\beta$)}
    \label{fig:percentage_ol_difficult}
\end{figure}

\begin{figure}[t]
    \centering
    \includegraphics[width=1\linewidth]{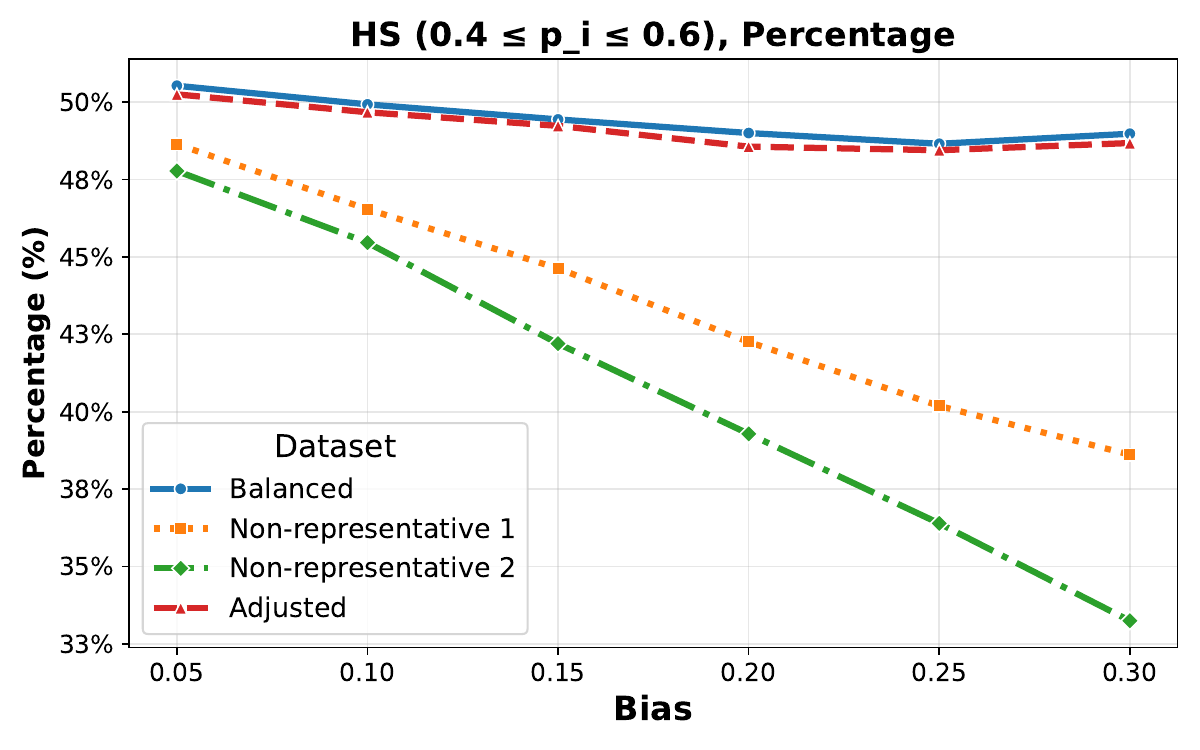}
    \caption{Percentage of HS instances on difficult tweets ($0.4 \leq p_{i,HS} \leq 0.6$) by dataset and bias ($\beta$)}
    \label{fig:percentage_hs_difficult}
\end{figure}

\paragraph{Accuracy.} Figure \ref{fig:f1-ol} compares the models' F1 scores. In contrast to Figure \ref{fig:acb-ol}, we do not see strong differences between the models trained on the different datasets. For all datasets, model performance declines with $\beta$: as the amount of bias in the annotations increases, the models are less able to predict the binary OL label. 

Because the F1 metric focuses on binary predictions, it is less sensitive to training biases than calibration metrics like ACB, which more explicitly capture biases through prediction scores. In decision-making, miscalibrated predictions can have harmful consequences when, for example, hateful content remains undetected \cite{van2019calibration}. These findings suggest that calibration metrics provide a clearer view of the impact of annotators on models: binary classification metrics can obscure such effects.

\subsection{HS Models}\label{app:hs}
Figure \ref{fig:acb-hs} contains the ACB results and Figure \ref{fig:f1-hs} the F1 score results for the HS models trained on each dataset. Though the adjusted model roughly tracks the representative models for ACB, there is instability in the results. All models show lower average ACB values than the representative model across a wide range of the bias offset (0.10 - 0.20). 
The PAIR approach does not improve calibration or accuracy: the adjusted model performs similarly to the Non-representative models. 
This effect is likely due to the combination of label rarity and our simulation design. With few positive annotations to begin with, the impact of the $\beta$ parameter and the overrepresentation of the A annotators may be overwhelmed by the baseline scarcity of hate speech annotations. Calibration metrics can be less reliable with rare classes \cite{calibration}.

\subsection{Sensitivity Analysis: Difficult Tweets}\label{app:difficult}
Our simulations assumed that all annotator type impacts all tweets the same way (Eq. \ref{eq:p_norm}), which is an oversimplification. More likely, \textbf{annotator characteristics have more impact for ambiguous tweets}. For example, prior research in the psychology literature on judgment under uncertainty  suggests that people draw more heavily on personal heuristics when interpreting unclear or underspecified information \cite{tversky1974judgment}. For this reason, we repeat model training and recompute metrics for those tweets where $0.4 \leq p_i \leq 0.6$. Subsetting the tweets in this way also eliminates the floor and ceiling effects in Eq. \ref{eq:p_norm}. The filtered datasets contain 267 (OL) and 360 (HS) tweets. The proportions of OL and HS annotations are stable for the Representative and Adjusted sets and decrease for the Non-representative sets as we increase the bias offset (Figures \ref{fig:percentage_ol_difficult} and \ref{fig:percentage_hs_difficult}). This mimics the trend for OL in the full dataset (Figure \ref{fig:percentage_ol}).

\begin{figure}[t]
    \centering
    \includegraphics[width=1\linewidth]{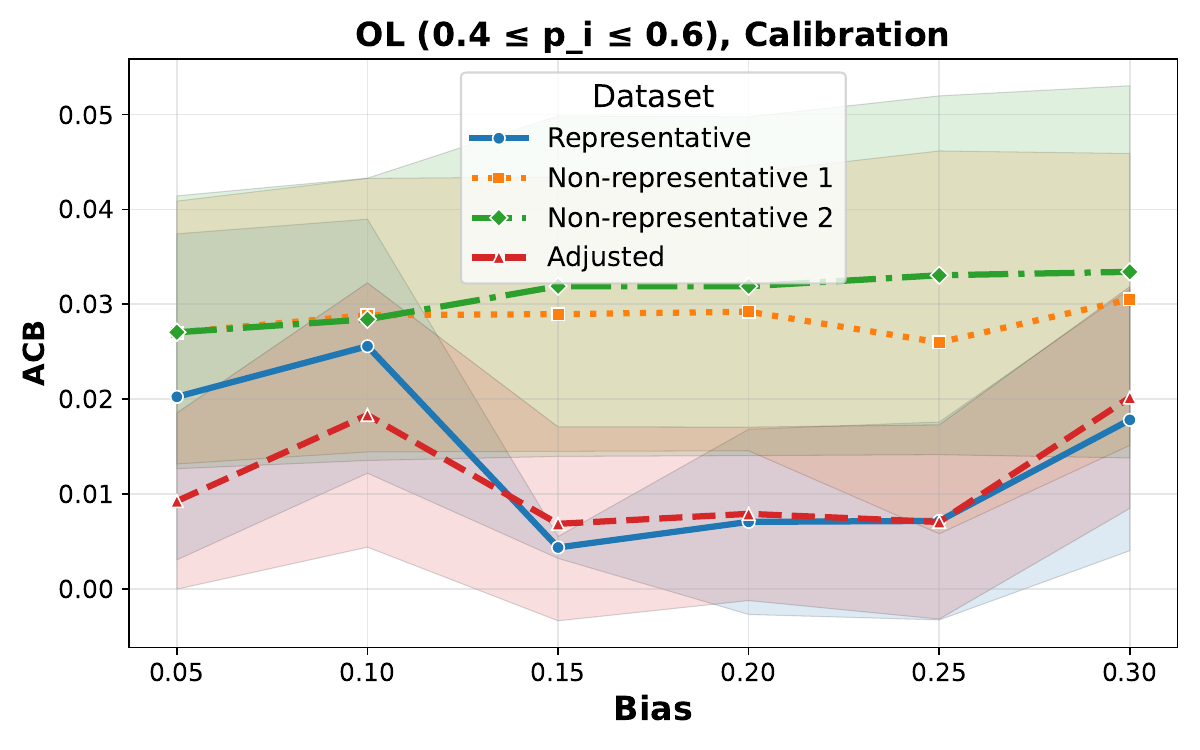}
    \caption{ACB scores for OL Models, on difficult tweets ($0.4 \leq p_{i,OL} \leq 0.6$), by dataset and bias ($\beta$)}
    \label{fig:acb-ol-filtered}
\end{figure}

\begin{figure}[t]
    \centering
    \includegraphics[width=1\linewidth]{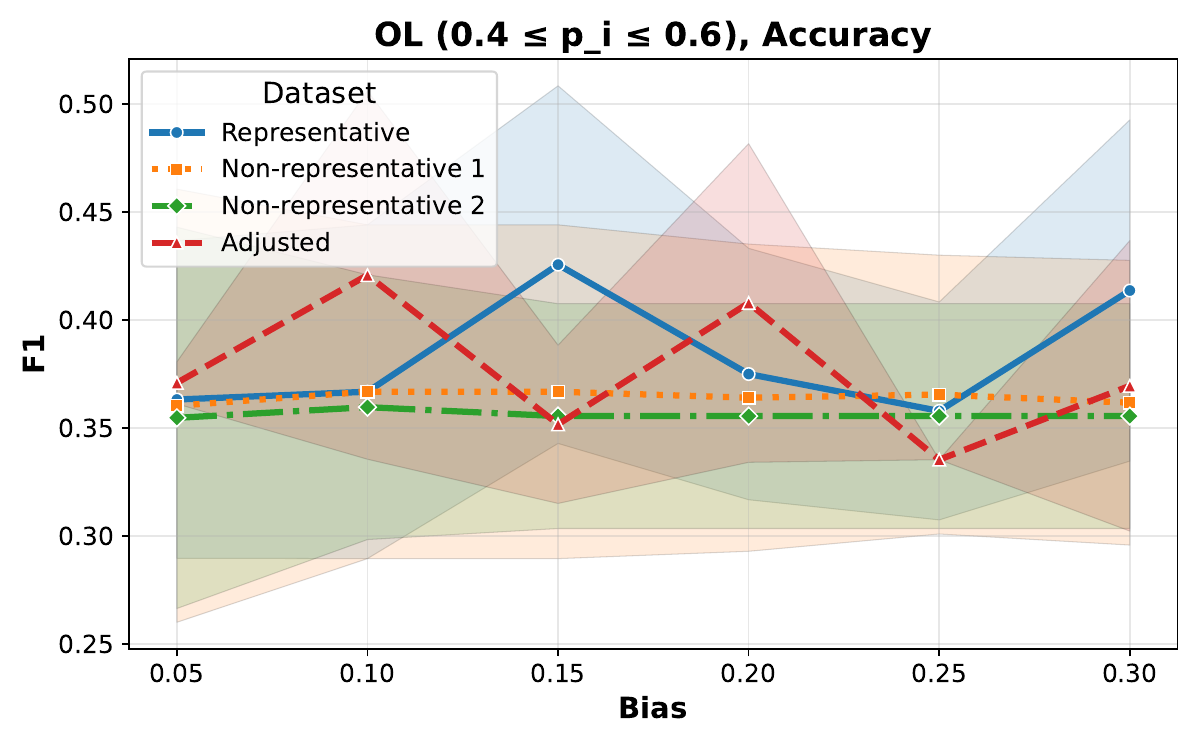}
    \caption{F1 scores for OL Models, on difficult tweets ($0.4 \leq p_{i,OL} \leq 0.6$), by dataset and bias ($\beta$)}
    \label{fig:f1-ol-filtered}
\end{figure}

\begin{figure}[t]
    \centering
    \includegraphics[width=1\linewidth]{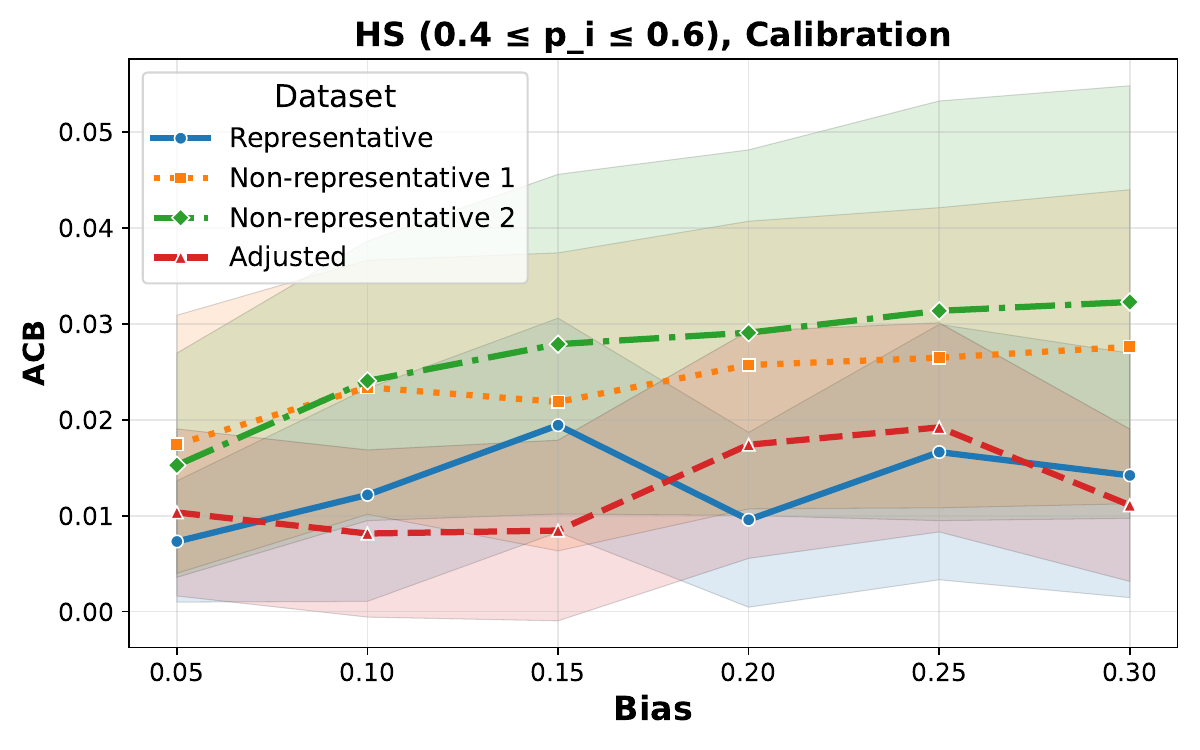}
    \caption{ACB scores for HS Models, on difficult tweets ($0.4 \leq p_{i,HS} \leq 0.6$), by dataset and bias ($\beta$)}
    \label{fig:acb-hs-filtered}
\end{figure}

\begin{figure}[t]
    \centering
    \includegraphics[width=1\linewidth]{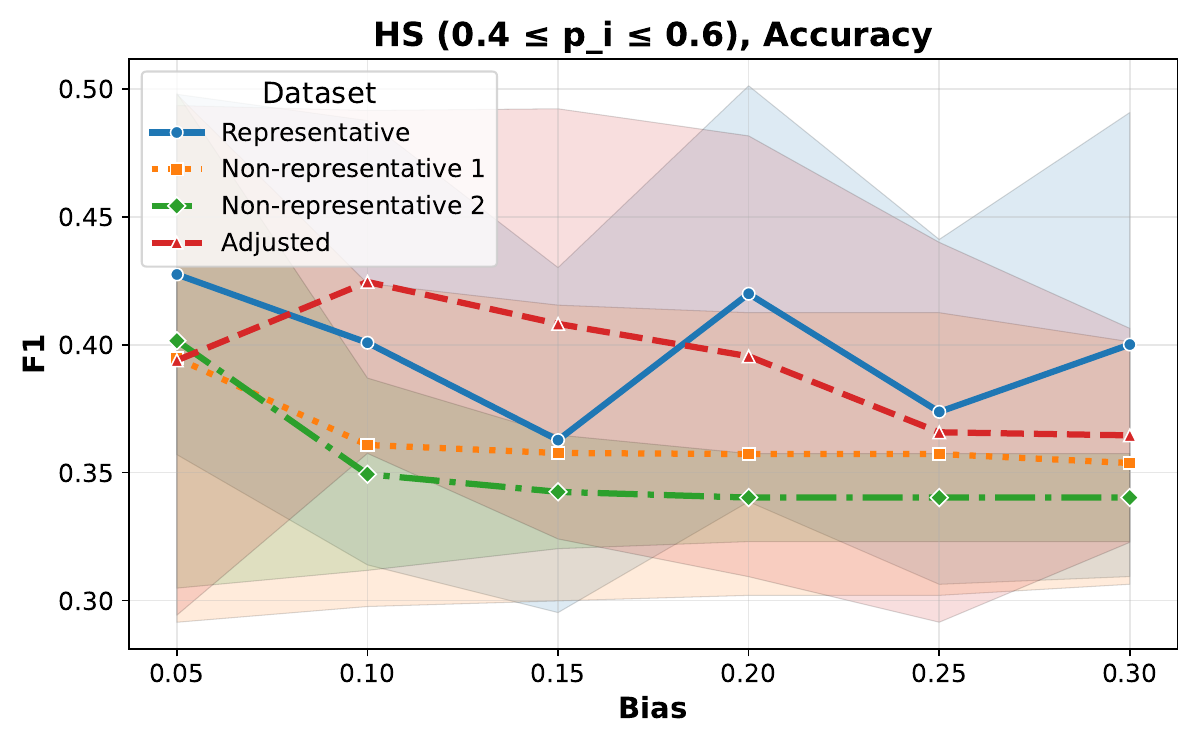}
    \caption{F1 scores for HS Models, on difficult tweets ($0.4 \leq p_{i,HS} \leq 0.6$), by dataset and bias ($\beta$)}
    \label{fig:f1-hs-filtered}
\end{figure}

Figures \ref{fig:acb-ol-filtered}, \ref{fig:f1-ol-filtered}, \ref{fig:acb-hs-filtered}, and \ref{fig:f1-hs-filtered} show results for two metrics (ACB, F1) for filtered OL and HS annotations. In Figure \ref{fig:acb-ol-filtered}, the Representative and Adjusted models have similar ACB and are lower than the Non-representative models. The F1 scores do not show differences between the models. These results are similar to those on the full set of tweets (Figures \ref{fig:acb-ol} and \ref{fig:f1-ol}). In the two HS figures (\ref{fig:acb-hs-filtered}, \ref{fig:f1-hs-filtered}), we see signs that the Representative and Adjusted models perform similarly, and better than the two Non-representative models, on both metrics. These results are more promising than those on the full set of tweets (Figures \ref{fig:acb-hs} and \ref{fig:f1-hs}) and support our hypothesis that the rarity of HS annotations contributed to the lack of positive results for the PAIR approach in \S\ref{app:hs}. The PAIR algorithm works well with difficult tweets, which is where it is likely most needed.

\section{Discussion \& Recommendations}

Our experimental results show the OL prediction models perform less well when trained on data from non-representative annotator pools (RQ1), and simple statistical adjustments can improve model calibration without collecting additional annotations or involving additional annotators (RQ2). While PAIR's impact was harder to assess for the rare HS class, PAIR did improve calibration of both the OL and HS models when trained on difficult tweets. These findings establish a promising bridge between survey statistics and machine learning -- offering a practical approach to make AI systems more representative of and responsive to the populations they serve, particularly for tasks involving subjective human judgments.

We recommend the following four steps to reduce bias due to non-representative annotator pools:

\begin{enumerate}[leftmargin=*]
\item[1)] \textbf{Use social science research} to identify the annotator characteristics that influence the propensity to engage in annotation and the annotations provided \cite{EckmanPosition}.

\item[2)] \textbf{Collect these characteristics} from annotators and gather corresponding population-level data from national censuses or high-quality surveys.\footnote{Collection and release of annotator characteristics or weights derived from them may raise confidentiality concerns. The survey literature offers advice for sharing sensitive data \citep[see][for a review]{Karr2016}. Collecting annotator characteristics may also require involvement of Institutional Review Boards or other participant protection organizations \cite{IRB}.}

\item[3)] \textbf{Calculate weights} that match the annotators to the population on those characteristics \citep[][]{Bethlehem2011, valliant2013practical}. 

\item[4)] \textbf{Use these weights in model training}. Our simple replication approach showed promise, and future work should test more sophisticated weighting approaches.
\end{enumerate}

\section{Limitations}
Our study explores bias-aware data simulation and evaluation in a controlled setting, which necessarily involves simplifying assumptions and methodological constraints. 
We outline key areas where future work could broaden the applicability and robustness of our findings.

\paragraph{Stylized Biases and Simulated Data.}
Our simulation makes strong assumptions about annotator behavior: there are only two types of annotators, and, within each type, annotators behave similarly. Real-world annotator biases may be more nuanced or context-dependent. The simulated annotators might not be representative of a stable opinion group \cite{mokhberian-etal-2024-capturing, vitsakis-etal-2024-voices}. 
Future work could incorporate more realistic biases and refine the proposed simulations and statistical techniques.

\paragraph{Sampling Variability.}
We have created only one version of the four datasets for each annotation type and value of $\beta$, each of which contains random draws from the Bernoulli distribution. A more traditional statistical approach would create multiple versions of the datasets and train models on each one, to average over the sampling variability. Though limited by computational constraints in this work, future work could take on a more expansive simulation. 
As discussed, we used five seeds in model training. 

\paragraph{Need for Population Benchmarks and Annotator Characteristics.} PAIR requires high quality benchmark information about the relevant population. These benchmarks might come from national statistical offices or national surveys. Annotators must provide accurate data on the same characteristics available in the benchmark data. Unfortunately, annotators sometimes do not provide accurate information \cite{Chandler2017, huang-etal-2023-incorporating}. In addition, theory demonstrates that bias will be reduced only when the characteristics used in weighting correlate with the annotations \citep{EckmanPosition}. In our simulation, differences in annotations were driven solely by group membership ($A$, $B$). In the real world, it is challenging to know what characteristics impact annotation behavior for a given task and to find good benchmarks for those characteristics.

\paragraph{Generalization Beyond Task Types.}
The study focuses only on binary classification tasks. Many real-world annotation tasks involve multiple classes or labels, which may show different bias patterns. Additional research is needed to extend these methods to more complex classification scenarios.

\paragraph{Evaluation Metrics.}
While we measured calibration and  accuracy, we did not examine other important metrics such as fairness across subgroups or robustness to adversarial examples. Future work on training data adjustment should assess a broader range of performance measures.

\section{Ethical Considerations}
In this simulation study, we experiment on a publicly available dataset collected in our previous study \cite{kern-etal-2023-annotation}, which contains offensive and hateful tweets. We do not support the views expressed in these tweets. 
The simulation study itself does not collect any new data or raise any ethical considerations.

\section*{Acknowledgments}
We acknowledge use of the Claude model to edit the text of the paper and to assist in coding. We thank the members of SODA Lab and MaiNLP labs from LMU Munich, and the members of the Social Data Science Group from University of Mannheim for their constructive feedback. This research is partially supported by RTI International, MCML, and BERD@NFDI. 
BP is supported by ERC Consolidator Grant DIALECT (101043235).

\bibliography{custom}

\begin{thebibliography}{51}
\providecommand{\natexlab}[1]{#1}

\bibitem[{Argyle et~al.(2023)Argyle, Busby, Fulda, Gubler, Rytting, and Wingate}]{argyle2023out}
Lisa~P Argyle, Ethan~C Busby, Nancy Fulda, Joshua~R Gubler, Christopher Rytting, and David Wingate. 2023.
\newblock \href {https://doi.org/10.1017/pan.2023.2} {Out of one, many: Using language models to simulate human samples}.
\newblock \emph{Political Analysis}, 31(3):339--355.

\bibitem[{Bender and Friedman(2018)}]{bender2018data}
Emily~M. Bender and Batya Friedman. 2018.
\newblock \href {https://doi.org/10.1162/tacl_a_00041} {Data statements for natural language processing: Toward mitigating system bias and enabling better science}.
\newblock \emph{Transactions of the Association for Computational Linguistics}, 6:587--604.

\bibitem[{Berinsky et~al.(2012)Berinsky, Huber, and Lenz}]{Berinsky_Huber_Lenz_2012}
Adam~J. Berinsky, Gregory~A. Huber, and Gabriel~S. Lenz. 2012.
\newblock \href {https://doi.org/10.1093/pan/mpr057} {Evaluating online labor markets for experimental research: Amazon.com’s mechanical turk}.
\newblock \emph{Political Analysis}, 20(3):351–368.

\bibitem[{Bethlehem et~al.(2011)Bethlehem, Cobben, and Schouten}]{Bethlehem2011}
Jelke Bethlehem, Fannie Cobben, and Barry Schouten. 2011.
\newblock \href {https://doi.org/10.1002/9780470891056} {\emph{Handbook of Nonresponse in Household Surveys}}.
\newblock Wiley.

\bibitem[{Burton et~al.(2006)Burton, Altman, Royston, and Holder}]{Burton2006}
Andrea Burton, Douglas~G. Altman, Patrick Royston, and Roger~L. Holder. 2006.
\newblock \href {https://doi.org/10.1002/sim.2673} {The design of simulation studies in medical statistics}.
\newblock \emph{Statistics in Medicine}, 25(24):4279–4292.

\bibitem[{Calders et~al.(2009)Calders, Kamiran, and Pechenizkiy}]{Caldersetal2009}
Toon Calders, Faisal Kamiran, and Mykola Pechenizkiy. 2009.
\newblock \href {https://doi.org/10.1109/ICDMW.2009.83} {Building classifiers with independency constraints}.
\newblock In \emph{ICDMW '09: Proceedings of the 2009 IEEE International Conference on Data Mining Workshops}, pages 13--18.

\bibitem[{Chandler and Paolacci(2017)}]{Chandler2017}
Jesse~J. Chandler and Gabriele Paolacci. 2017.
\newblock \href {https://doi.org/10.1177/1948550617698203} {Lie for a dime: When most prescreening responses are honest but most study participants are impostors}.
\newblock \emph{Social Psychological and Personality Science}, 8(5):500–508.

\bibitem[{Eckman et~al.(2024)Eckman, Plank, and Kreuter}]{EckmanPosition}
Stephanie Eckman, Barbara Plank, and Frauke Kreuter. 2024.
\newblock \href {https://proceedings.mlr.press/v235/eckman24a.html} {Position: Insights from survey methodology can improve training data}.
\newblock In \emph{Proceedings of the 41st International Conference on Machine Learning}, volume 235 of \emph{Proceedings of Machine Learning Research}, pages 12268--12283. PMLR.

\bibitem[{Favier et~al.(2023)Favier, Calders, Pinxteren, and Meyer}]{favier2023fair}
Marco Favier, Toon Calders, Sam Pinxteren, and Jonathan Meyer. 2023.
\newblock \href {https://doi.org/10.1007/s10994-023-06401-1} {How to be fair? a study of label and selection bias}.
\newblock \emph{Machine Learning}, 112(12):5081--5104.

\bibitem[{Fleisig et~al.(2023)Fleisig, Abebe, and Klein}]{Fleisig2023WhenTM}
Eve Fleisig, Rediet Abebe, and Dan Klein. 2023.
\newblock \href {https://doi.org/10.18653/v1/2023.emnlp-main.415} {When the majority is wrong: Modeling annotator disagreement for subjective tasks}.
\newblock In \emph{Proceedings of the 2023 Conference on Empirical Methods in Natural Language Processing}, pages 6715--6726, Singapore. Association for Computational Linguistics.

\bibitem[{Fleisig et~al.(2024)Fleisig, Blodgett, Klein, and Talat}]{Fleisigetal05}
Eve Fleisig, Su~Lin Blodgett, Dan Klein, and Zeerak Talat. 2024.
\newblock \href {https://doi.org/10.18653/v1/2024.naacl-long.126} {The perspectivist paradigm shift: Assumptions and challenges of capturing human labels}.
\newblock In \emph{Proceedings of the 2024 Conference of the North American Chapter of the Association for Computational Linguistics: Human Language Technologies (Volume 1: Long Papers)}, pages 2279--2292, Mexico City, Mexico. Association for Computational Linguistics.

\bibitem[{Giorgi et~al.(2025)Giorgi, Cima, Fagni, Avvenuti, and Cresci}]{giorgi2024human}
Tommaso Giorgi, Lorenzo Cima, Tiziano Fagni, Marco Avvenuti, and Stefano Cresci. 2025.
\newblock \href {https://doi.org/10.1609/icwsm.v19i1.35837} {Human and llm biases in hate speech annotations: A socio-demographic analysis of annotators and targets}.
\newblock \emph{Proceedings of the International AAAI Conference on Web and Social Media}, 19(1):653--670.

\bibitem[{Hebert-Johnson et~al.(2018)Hebert-Johnson, Kim, Reingold, and Rothblum}]{multicalibration}
Ursula Hebert-Johnson, Michael Kim, Omer Reingold, and Guy Rothblum. 2018.
\newblock \href {https://proceedings.mlr.press/v80/hebert-johnson18a.html} {Multicalibration: Calibration for the ({C}omputationally-identifiable) masses}.
\newblock In \emph{Proceedings of the 35th International Conference on Machine Learning}, volume~80 of \emph{Proceedings of Machine Learning Research}, pages 1939--1948. PMLR.

\bibitem[{Huang et~al.(2023)Huang, Fleisig, and Klein}]{huang-etal-2023-incorporating}
Olivia Huang, Eve Fleisig, and Dan Klein. 2023.
\newblock \href {https://doi.org/10.18653/v1/2023.emnlp-main.64} {Incorporating worker perspectives into {MT}urk annotation practices for {NLP}}.
\newblock In \emph{Proceedings of the 2023 Conference on Empirical Methods in Natural Language Processing}, pages 1010--1028, Singapore. Association for Computational Linguistics.

\bibitem[{H{\"u}llermeier and Waegeman(2021)}]{hullermeier2021aleatoric}
Eyke H{\"u}llermeier and Willem Waegeman. 2021.
\newblock \href {https://doi.org/10.1007/s10994-021-05946-3} {Aleatoric and epistemic uncertainty in machine learning: An introduction to concepts and methods}.
\newblock \emph{Machine learning}, 110(3):457--506.

\bibitem[{Kamiran and Calders(2012)}]{kamiran2012data}
Faisal Kamiran and Toon Calders. 2012.
\newblock \href {https://doi.org/10.1007/s10115-011-0463-8} {Data preprocessing techniques for classification without discrimination}.
\newblock \emph{Knowledge and Information Systems}, 33(1):1--33.

\bibitem[{Karr(2016)}]{Karr2016}
Alan~F. Karr. 2016.
\newblock \href {https://doi.org/10.1146/annurev-statistics-041715-033438} {Data sharing and access}.
\newblock \emph{Annual Review of Statistics and Its Application}, 3(Volume 3, 2016):113--132.

\bibitem[{Kaushik et~al.(2024)Kaushik, Lipton, and London}]{IRB}
Divyansh Kaushik, Zachary~C. Lipton, and Alex~John London. 2024.
\newblock \href {https://doi.org/10.1145/3641858} {Resolving the human-subjects status of ml's crowdworkers}.
\newblock \emph{Commun. ACM}, 67(5):52–59.

\bibitem[{Kern et~al.(2023)Kern, Eckman, Beck, Chew, Ma, and Kreuter}]{kern-etal-2023-annotation}
Christoph Kern, Stephanie Eckman, Jacob Beck, Rob Chew, Bolei Ma, and Frauke Kreuter. 2023.
\newblock \href {https://doi.org/10.18653/v1/2023.findings-emnlp.992} {Annotation sensitivity: Training data collection methods affect model performance}.
\newblock In \emph{Findings of the Association for Computational Linguistics: EMNLP 2023}, pages 14874--14886, Singapore. Association for Computational Linguistics.

\bibitem[{Kirk et~al.(2024)Kirk, Whitefield, R\"{o}ttger, Bean, Margatina, Ciro, Mosquera, Bartolo, Williams, He, Vidgen, and Hale}]{Kirk_PRISM}
Hannah~Rose Kirk, Alexander Whitefield, Paul R\"{o}ttger, Andrew Bean, Katerina Margatina, Juan Ciro, Rafael Mosquera, Max Bartolo, Adina Williams, He~He, Bertie Vidgen, and Scott~A. Hale. 2024.
\newblock \href {https://proceedings.neurips.cc/paper_files/paper/2024/file/be2e1b68b44f2419e19f6c35a1b8cf35-Paper-Datasets_and_Benchmarks_Track.pdf} {The prism alignment dataset: What participatory, representative and individualised human feedback reveals about the subjective and multicultural alignment of large language models}.
\newblock In \emph{Advances in Neural Information Processing Systems}, volume~37, pages 105236--105344. Curran Associates, Inc.

\bibitem[{Leonardelli et~al.(2023)Leonardelli, Abercrombie, Almanea, Basile, Fornaciari, Plank, Rieser, Uma, and Poesio}]{leonardelli2023semeval2023task11learning}
Elisa Leonardelli, Gavin Abercrombie, Dina Almanea, Valerio Basile, Tommaso Fornaciari, Barbara Plank, Verena Rieser, Alexandra Uma, and Massimo Poesio. 2023.
\newblock \href {https://doi.org/10.18653/v1/2023.semeval-1.314} {{S}em{E}val-2023 task 11: Learning with disagreements ({L}e{W}i{D}i)}.
\newblock In \emph{Proceedings of the 17th International Workshop on Semantic Evaluation (SemEval-2023)}, pages 2304--2318, Toronto, Canada. Association for Computational Linguistics.

\bibitem[{Ling and Li(1998)}]{LingDirectMarketing}
Charles~X. Ling and Chenghui Li. 1998.
\newblock \href {https://cdn.aaai.org/KDD/1998/KDD98-011.pdf} {Data mining for direct marketing: problems and solutions}.
\newblock In \emph{Proceedings of the Fourth International Conference on Knowledge Discovery and Data Mining}, KDD'98, page 73–79. AAAI Press.

\bibitem[{Liu et~al.(2019)Liu, Ott, Goyal, Du, Joshi, Chen, Levy, Lewis, Zettlemoyer, and Stoyanov}]{liu2019roberta}
Yinhan Liu, Myle Ott, Naman Goyal, Jingfei Du, Mandar Joshi, Danqi Chen, Omer Levy, Mike Lewis, Luke Zettlemoyer, and Veselin Stoyanov. 2019.
\newblock \href {https://arxiv.org/abs/1907.11692} {Roberta: A robustly optimized bert pretraining approach}.
\newblock \emph{Preprint}, arXiv:1907.11692.

\bibitem[{Lowmanstone et~al.(2023)Lowmanstone, Wan, Owan, Kim, and Kang}]{LowmanstoneWOKK23}
London Lowmanstone, Ruyuan Wan, Risako Owan, Jaehyung Kim, and Dongyeop Kang. 2023.
\newblock \href {https://ceur-ws.org/Vol-3494/paper10.pdf} {Annotation imputation to individualize predictions: Initial studies on distribution dynamics and model predictions}.
\newblock In \emph{NLPerspectives@ECAI}.

\bibitem[{Mehrabi et~al.(2021)Mehrabi, Morstatter, Saxena, Lerman, and Galstyan}]{mehrabi_survey_2021}
Ninareh Mehrabi, Fred Morstatter, Nripsuta Saxena, Kristina Lerman, and Aram Galstyan. 2021.
\newblock \href {https://doi.org/10.1145/3457607} {{A Survey on Bias and Fairness in Machine Learning}}.
\newblock \emph{ACM Computing Surveys}, 54(6):1--36.

\bibitem[{Mokhberian et~al.(2024)Mokhberian, Marmarelis, Hopp, Basile, Morstatter, and Lerman}]{mokhberian-etal-2024-capturing}
Negar Mokhberian, Myrl Marmarelis, Frederic Hopp, Valerio Basile, Fred Morstatter, and Kristina Lerman. 2024.
\newblock \href {https://doi.org/10.18653/v1/2024.naacl-long.407} {Capturing perspectives of crowdsourced annotators in subjective learning tasks}.
\newblock In \emph{Proceedings of the 2024 Conference of the North American Chapter of the Association for Computational Linguistics: Human Language Technologies (Volume 1: Long Papers)}, pages 7337--7349, Mexico City, Mexico. Association for Computational Linguistics.

\bibitem[{Morris et~al.(2019)Morris, White, and Crowther}]{Morris2019}
Tim~P. Morris, Ian~R. White, and Michael~J. Crowther. 2019.
\newblock \href {https://doi.org/10.1002/sim.8086} {Using simulation studies to evaluate statistical methods}.
\newblock \emph{Statistics in Medicine}, 38(11):2074–2102.

\bibitem[{Naeini et~al.(2015)Naeini, Cooper, and Hauskrecht}]{Naeini_Cooper_Hauskrecht_2015}
Mahdi~Pakdaman Naeini, Gregory Cooper, and Milos Hauskrecht. 2015.
\newblock \href {https://doi.org/10.1609/aaai.v29i1.9602} {Obtaining well calibrated probabilities using bayesian binning}.
\newblock \emph{Proceedings of the AAAI Conference on Artificial Intelligence}, 29(1).

\bibitem[{Orlikowski et~al.(2025)Orlikowski, Pei, R{\"o}ttger, Cimiano, Jurgens, and Hovy}]{orlikowski2025demographics}
Matthias Orlikowski, Jiaxin Pei, Paul R{\"o}ttger, Philipp Cimiano, David Jurgens, and Dirk Hovy. 2025.
\newblock \href {https://aclanthology.org/2025.acl-long.104/} {Beyond demographics: Fine-tuning large language models to predict individuals' subjective text perceptions}.
\newblock In \emph{Proceedings of the 63rd Annual Meeting of the Association for Computational Linguistics (Volume 1: Long Papers)}, pages 2092--2111, Vienna, Austria. Association for Computational Linguistics.

\bibitem[{Ouyang et~al.(2022)Ouyang, Wu, Jiang, Almeida, Wainwright, Mishkin, Zhang, Agarwal, Slama, Ray, Schulman, Hilton, Kelton, Miller, Simens, Askell, Welinder, Christiano, Leike, and Lowe}]{ouyang2022traininglanguagemodelsfollow}
Long Ouyang, Jeff Wu, Xu~Jiang, Diogo Almeida, Carroll~L. Wainwright, Pamela Mishkin, Chong Zhang, Sandhini Agarwal, Katarina Slama, Alex Ray, John Schulman, Jacob Hilton, Fraser Kelton, Luke Miller, Maddie Simens, Amanda Askell, Peter Welinder, Paul Christiano, Jan Leike, and Ryan Lowe. 2022.
\newblock \href {https://arxiv.org/abs/2203.02155} {Training language models to follow instructions with human feedback}.
\newblock \emph{Preprint}, arXiv:2203.02155.

\bibitem[{Paszke et~al.(2019)Paszke, Gross, Massa, Lerer, Bradbury, Chanan, Killeen, Lin, Gimelshein, Antiga, Desmaison, Kopf, Yang, DeVito, Raison, Tejani, Chilamkurthy, Steiner, Fang, Bai, and Chintala}]{pytorch2019}
Adam Paszke, Sam Gross, Francisco Massa, Adam Lerer, James Bradbury, Gregory Chanan, Trevor Killeen, Zeming Lin, Natalia Gimelshein, Luca Antiga, Alban Desmaison, Andreas Kopf, Edward Yang, Zachary DeVito, Martin Raison, Alykhan Tejani, Sasank Chilamkurthy, Benoit Steiner, Lu~Fang, Junjie Bai, and Soumith Chintala. 2019.
\newblock \href {http://papers.neurips.cc/paper/9015-pytorch-an-imperative-style-high-performance-deep-learning-library.pdf} {Pytorch: An imperative style, high-performance deep learning library}.
\newblock In \emph{Advances in Neural Information Processing Systems 32}, pages 8024--8035. Curran Associates, Inc.

\bibitem[{Pei and Jurgens(2023)}]{pei-jurgens-2023-annotator}
Jiaxin Pei and David Jurgens. 2023.
\newblock \href {https://doi.org/10.18653/v1/2023.law-1.25} {When do annotator demographics matter? measuring the influence of annotator demographics with the {POPQUORN} dataset}.
\newblock In \emph{Proceedings of the 17th Linguistic Annotation Workshop (LAW-XVII)}, pages 252--265, Toronto, Canada. Association for Computational Linguistics.

\bibitem[{Plank(2022)}]{plank-2022-problem}
Barbara Plank. 2022.
\newblock \href {https://doi.org/10.18653/v1/2022.emnlp-main.731} {The ``problem'' of human label variation: On ground truth in data, modeling and evaluation}.
\newblock In \emph{Proceedings of the 2022 Conference on Empirical Methods in Natural Language Processing}, pages 10671--10682, Abu Dhabi, United Arab Emirates. Association for Computational Linguistics.

\bibitem[{Prabhakaran et~al.(2021)Prabhakaran, Mostafazadeh~Davani, and Diaz}]{Prabhakaran2021}
Vinodkumar Prabhakaran, Aida Mostafazadeh~Davani, and Mark Diaz. 2021.
\newblock \href {https://doi.org/10.18653/v1/2021.law-1.14} {On releasing annotator-level labels and information in datasets}.
\newblock In \emph{Proceedings of the Joint 15th Linguistic Annotation Workshop (LAW) and 3rd Designing Meaning Representations (DMR) Workshop}, pages 133--138, Punta Cana, Dominican Republic. Association for Computational Linguistics.

\bibitem[{Quatember(2015)}]{quatember-2015}
Andreas Quatember. 2015.
\newblock \href {https://doi.org/10.1007/978-3-319-11785-0} {\emph{Pseudo-Populations: A Basic Concept in Statistical Surveys}}.
\newblock Springer.

\bibitem[{Roelofs et~al.(2022)Roelofs, Cain, Shlens, and Mozer}]{pmlr-v151-roelofs22a}
Rebecca Roelofs, Nicholas Cain, Jonathon Shlens, and Michael~C. Mozer. 2022.
\newblock \href {https://proceedings.mlr.press/v151/roelofs22a.html} {Mitigating bias in calibration error estimation}.
\newblock In \emph{Proceedings of The 25th International Conference on Artificial Intelligence and Statistics}, volume 151 of \emph{Proceedings of Machine Learning Research}, pages 4036--4054. PMLR.

\bibitem[{Rolf et~al.(2021)Rolf, Worledge, Recht, and Jordan}]{rolf2021representation}
Esther Rolf, Theodora~T Worledge, Benjamin Recht, and Michael Jordan. 2021.
\newblock \href {https://proceedings.mlr.press/v139/rolf21a.html} {Representation matters: Assessing the importance of subgroup allocations in training data}.
\newblock In \emph{Proceedings of the 38th International Conference on Machine Learning}, volume 139 of \emph{Proceedings of Machine Learning Research}, pages 9040--9051. PMLR.

\bibitem[{Santy et~al.(2023)Santy, Liang, Le~Bras, Reinecke, and Sap}]{santy-etal-2023-nlpositionality}
Sebastin Santy, Jenny Liang, Ronan Le~Bras, Katharina Reinecke, and Maarten Sap. 2023.
\newblock \href {https://doi.org/10.18653/v1/2023.acl-long.505} {{NLP}ositionality: Characterizing design biases of datasets and models}.
\newblock In \emph{Proceedings of the 61st Annual Meeting of the Association for Computational Linguistics (Volume 1: Long Papers)}, pages 9080--9102, Toronto, Canada. Association for Computational Linguistics.

\bibitem[{Sap et~al.(2022)Sap, Swayamdipta, Vianna, Zhou, Choi, and Smith}]{sap2022annotators}
Maarten Sap, Swabha Swayamdipta, Laura Vianna, Xuhui Zhou, Yejin Choi, and Noah~A. Smith. 2022.
\newblock \href {https://doi.org/10.18653/v1/2022.naacl-main.431} {Annotators with attitudes: How annotator beliefs and identities bias toxic language detection}.
\newblock In \emph{Proceedings of the 2022 Conference of the North American Chapter of the Association for Computational Linguistics: Human Language Technologies}, pages 5884--5906, Seattle, United States. Association for Computational Linguistics.

\bibitem[{Smart et~al.(2024)Smart, Wang, Monk, Díaz, Kasirzadeh, Liemt, and Schmer-Galunder}]{smart2024disciplinelabelweirdgenealogy}
Andrew Smart, Ding Wang, Ellis Monk, Mark Díaz, Atoosa Kasirzadeh, Erin~Van Liemt, and Sonja Schmer-Galunder. 2024.
\newblock \href {https://arxiv.org/abs/2402.06811} {Discipline and label: A weird genealogy and social theory of data annotation}.
\newblock \emph{Preprint}, arXiv:2402.06811.

\bibitem[{Sorensen et~al.(2024)Sorensen, Moore, Fisher, Gordon, Mireshghallah, Rytting, Ye, Jiang, Lu, Dziri, Althoff, and Choi}]{sorensen2024position}
Taylor Sorensen, Jared Moore, Jillian Fisher, Mitchell~L Gordon, Niloofar Mireshghallah, Christopher~Michael Rytting, Andre Ye, Liwei Jiang, Ximing Lu, Nouha Dziri, Tim Althoff, and Yejin Choi. 2024.
\newblock \href {https://proceedings.mlr.press/v235/sorensen24a.html} {Position: A roadmap to pluralistic alignment}.
\newblock In \emph{Proceedings of the 41st International Conference on Machine Learning}, volume 235 of \emph{Proceedings of Machine Learning Research}, pages 46280--46302. PMLR.

\bibitem[{Sun et~al.(2025)Sun, Pei, Choi, and Jurgens}]{sun-etal-2025-sociodemographic}
Huaman Sun, Jiaxin Pei, Minje Choi, and David Jurgens. 2025.
\newblock \href {https://aclanthology.org/2025.naacl-short.71/} {Sociodemographic prompting is not yet an effective approach for simulating subjective judgments with {LLM}s}.
\newblock In \emph{Proceedings of the 2025 Conference of the Nations of the Americas Chapter of the Association for Computational Linguistics: Human Language Technologies (Volume 2: Short Papers)}, pages 845--854, Albuquerque, New Mexico. Association for Computational Linguistics.

\bibitem[{Tversky and Kahneman(1974)}]{tversky1974judgment}
Amos Tversky and Daniel Kahneman. 1974.
\newblock \href {https://doi.org/10.1126/science.185.4157.1124} {Judgment under uncertainty: Heuristics and biases: Biases in judgments reveal some heuristics of thinking under uncertainty.}
\newblock \emph{science}, 185(4157):1124--1131.

\bibitem[{Uma et~al.(2021)Uma, Fornaciari, Hovy, Paun, Plank, and Poesio}]{uma2021learning}
Alexandra~N Uma, Tommaso Fornaciari, Dirk Hovy, Silviu Paun, Barbara Plank, and Massimo Poesio. 2021.
\newblock \href {https://doi.org/10.1613/jair.1.12752} {Learning from disagreement: A survey}.
\newblock \emph{Journal of Artificial Intelligence Research}, 72:1385--1470.

\bibitem[{Valliant(2019)}]{Valliant2019}
Richard Valliant. 2019.
\newblock \href {https://doi.org/10.1093/jssam/smz003} {Comparing alternatives for estimation from nonprobability samples}.
\newblock \emph{Journal of Survey Statistics and Methodology}, 8(2):231–263.

\bibitem[{Valliant et~al.(2013)Valliant, Dever, and Kreuter}]{valliant2013practical}
Richard Valliant, Jill~A Dever, and Frauke Kreuter. 2013.
\newblock \href {https://doi.org/10.1007/978-3-319-93632-1} {\emph{Practical tools for designing and weighting survey samples}}, volume~1.
\newblock Springer.

\bibitem[{Van~Calster et~al.(2019)Van~Calster, McLernon, van Smeden, Wynants, Steyerberg, Bossuyt, Collins, Macaskill, McLernon, Moons, Steyerberg, Van~Calster, van Smeden, and Vickers}]{van2019calibration}
Ben Van~Calster, David~J. McLernon, Maarten van Smeden, Laure Wynants, Ewout~W. Steyerberg, Patrick Bossuyt, Gary~S. Collins, Petra Macaskill, David~J. McLernon, Karel G.~M. Moons, Ewout~W. Steyerberg, Ben Van~Calster, Maarten van Smeden, and Andrew~J. Vickers. 2019.
\newblock \href {https://doi.org/10.1186/s12916-019-1466-7} {Calibration: the achilles heel of predictive analytics}.
\newblock \emph{BMC Medicine}, 17(1):230.

\bibitem[{Vitsakis et~al.(2024)Vitsakis, Parekh, and Konstas}]{vitsakis-etal-2024-voices}
Nikolas Vitsakis, Amit Parekh, and Ioannis Konstas. 2024.
\newblock \href {https://doi.org/10.18653/v1/2024.emnlp-main.696} {Voices in a crowd: Searching for clusters of unique perspectives}.
\newblock In \emph{Proceedings of the 2024 Conference on Empirical Methods in Natural Language Processing}, pages 12517--12539, Miami, Florida, USA. Association for Computational Linguistics.

\bibitem[{Wan et~al.(2023)Wan, Kim, and Kang}]{Wan2023}
Ruyuan Wan, Jaehyung Kim, and Dongyeop Kang. 2023.
\newblock \href {https://doi.org/10.1609/aaai.v37i12.26698} {Everyone’s voice matters: Quantifying annotation disagreement using demographic information}.
\newblock \emph{Proceedings of the AAAI Conference on Artificial Intelligence}, 37(12):14523--14530.

\bibitem[{Wolf et~al.(2020)Wolf, Debut, Sanh, Chaumond, Delangue, Moi, Cistac, Rault, Louf, Funtowicz, Davison, Shleifer, von Platen, Ma, Jernite, Plu, Xu, Le~Scao, Gugger, Drame, Lhoest, and Rush}]{wolf-etal-2020-transformers}
Thomas Wolf, Lysandre Debut, Victor Sanh, Julien Chaumond, Clement Delangue, Anthony Moi, Pierric Cistac, Tim Rault, Remi Louf, Morgan Funtowicz, Joe Davison, Sam Shleifer, Patrick von Platen, Clara Ma, Yacine Jernite, Julien Plu, Canwen Xu, Teven Le~Scao, Sylvain Gugger, Mariama Drame, Quentin Lhoest, and Alexander Rush. 2020.
\newblock \href {https://doi.org/10.18653/v1/2020.emnlp-demos.6} {Transformers: State-of-the-art natural language processing}.
\newblock In \emph{Proceedings of the 2020 Conference on Empirical Methods in Natural Language Processing: System Demonstrations}, pages 38--45, Online. Association for Computational Linguistics.

\bibitem[{Zhong et~al.(2021)Zhong, Cui, Liu, and Jia}]{calibration}
Zhisheng Zhong, Jiequan Cui, Shu Liu, and Jiaya Jia. 2021.
\newblock \href {https://doi.org/10.1109/CVPR46437.2021.01622} {Improving calibration for long-tailed recognition}.
\newblock In \emph{2021 IEEE/CVF Conference on Computer Vision and Pattern Recognition (CVPR)}, pages 16484--16493.

\end{thebibliography}

\end{document}